\documentclass[conference]{IEEEtran}
\IEEEoverridecommandlockouts
\usepackage{amsmath,amssymb,amsfonts}
\usepackage{algorithmic}
\usepackage{graphicx}
\usepackage{textcomp}
\usepackage{xcolor}
\usepackage{orcidlink}

\usepackage[backend=biber, style=numeric]{biblatex}

\def\BibTeX{{\rm B\kern-.05em{\sc i\kern-.025em b}\kern-.08em
    T\kern-.1667em\lower.7ex\hbox{E}\kern-.125emX}}
\begin{document}

\title{Identification and Mitigating Bias in Quantum Machine Learning \\}

\author{\IEEEauthorblockN{Nandhini Swaminathan \orcidlink{https://orcid.org/0000-0002-9835-6704}}
\IEEEauthorblockA{\textit{Computer Science and Engineering} \\
\textit{University of California, San Diego}\\
La Jolla, San Diego \\
}
\and
\IEEEauthorblockN{David Danks\orcidlink{https://orcid.org/0000-0003-4541-5966}}
\IEEEauthorblockA{\textit{Data Science, Philosophy, and Policy} \\
\textit{University of California, San Diego}\\
La Jolla, San Diego \\}
}

\maketitle

\begin{abstract}
As quantum machine learning (QML) emerges as a promising field at the intersection of quantum computing and artificial intelligence, it becomes crucial to address the biases and challenges that arise from the unique nature of quantum systems.  This research includes work on identification, diagnosis, and response to biases in Quantum Machine Learning. This paper aims to provide an overview of three key topics: How does bias unique to Quantum Machine Learning look?  Why and how can it occur? What can and should be done about it? 
\end{abstract}


\section{Introduction}In classical machine learning, the importance of fairness and bias mitigation has become increasingly recognized, leading to significant research efforts and practical implementations. As Quantum Machine Learning (QML) rapidly advances and finds applications in critical domains such as drug discovery, finance, and cryptography, it becomes imperative to extend these fairness considerations to the quantum realm.
While recent work has begun to address fairness in QML (e.g., development of interpretable models like Q-LIME and Q-Shapley), a comprehensive examination of biases specific to quantum systems has yet to be established. This research aims to fill that gap with these key contributions:
\begin{enumerate}
    \item We identify and analyze several critical instances of bias in QML systems, including the sources of these biases (i.e., data, algorithm, and  measurement) and their implications for QML model performance and fairness. 
    \item We present quantitative simulation results demonstrating the tangible impact of encoding bias on QML outcomes.
    \item We provide an overview of current mitigation strategies for these biases, drawing from both classical and quantum techniques.
\end{enumerate}

\section{Sources \& Types of Biases}
We propose that quantum-specific biases in QML can arise through features of the data, algorithms, or measurements and consider each of them.

\subsection*{Data Representation Bias}
\begin{enumerate}
\item \textbf{Encoding Bias}\textit{ arises from the interaction between the transformation of classical data into quantum states,
and the quantum algorithm. } 
\end{enumerate}

 To examine this bias, we conducted an experiment\footnote{ Experiment code available at:   https://github.com/nandhiniswaminathan/QML-Encoding-Methods/tree/main} using various encoding techniques applied to a fixed QNN architecture on the MNIST dataset for classification tasks. Our results (see Figure. 1)  show the significant impact of encoding choice on model performance:

\begin{itemize}
    \item Basis Encoding exhibited consistently low accuracy across all epochs, indicating limited learning capacity with this encoding method.
    \item Angle Encoding demonstrated rapid improvement in accuracy, quickly reaching and maintaining high performance levels after just a few epochs.
    \item Hybrid Parameterized Encoding, tested with different rotation axes ($R_x$, $R_y$, $R_z$), showed varying behaviors:
    \begin{itemize}
        \item $R_x$ encoding exhibited fluctuating performance with an overall upward trend.
        \item $R_y$ encoding showed the most rapid initial improvement followed by consistently high accuracy.
        \item $R_z$ encoding initially maintained poor accuracy and experienced a significant drop before recovering in later epochs.
    \end{itemize}
\end{itemize}
Our experiment demonstrates Encoding bias in QML. We observe that angle and $R_y$ encoding were the most robust. Current approaches to mitigate this bias  involve comprehensive simulation and analysis of various encoding strategies and algorithmic combinations, as demonstrated in our experiment.

\begin{figure}
    \centering
    \includegraphics[width=0.5\textwidth]{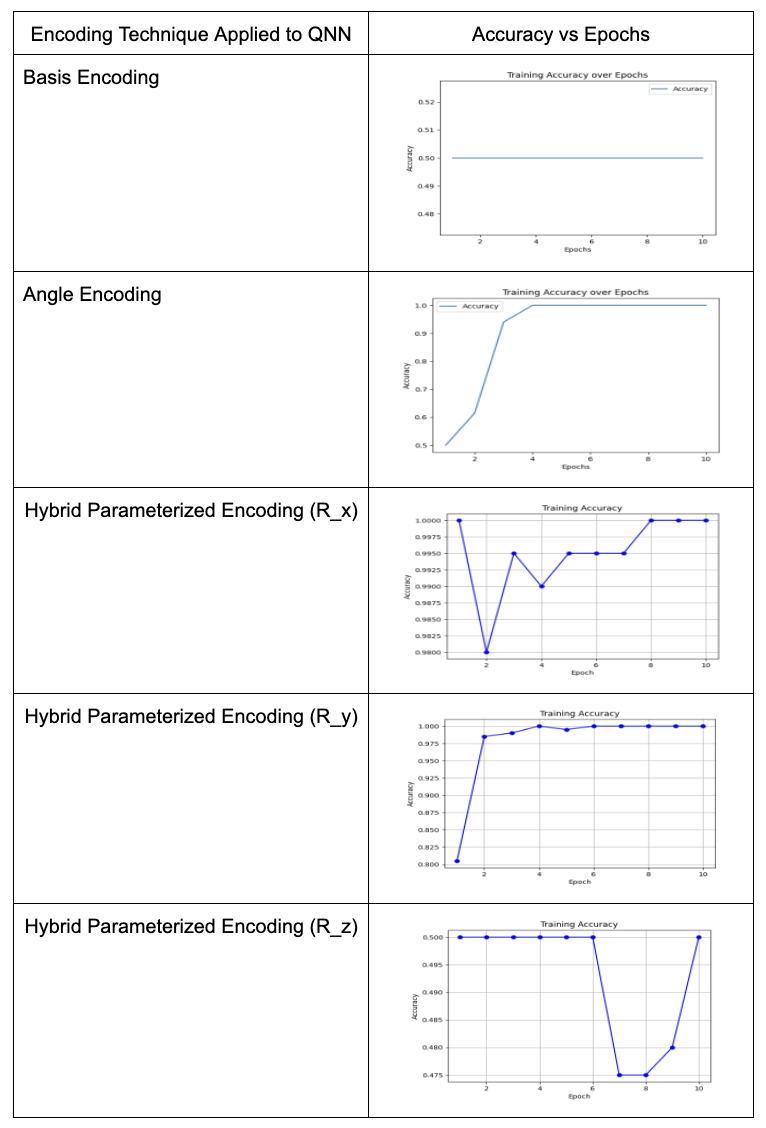}
\caption{Performance differences on MNIST dataset due to Encoding Bias}
    \label{fig:encoding}
\end{figure}

\subsection*{Algorithmic Biases}
\begin{enumerate}
\item \textbf{Inductive Bias}\textit{ is due to the assumptions required for an algorithm to appropriately model or predict previously unseen scenarios.}
As the number of qubits \textit{n} increases, the dimension of the Hilbert space grows exponentially, and random quantum states tend to be almost orthogonal to each other due to the ``concentration of measure'' phenomenon. Kuebler et al \cite{kubler2021inductive} leverage this to demonstrate that the largest eigenvalue of a kernel matrix constructed from the inner products of these quantum states is extremely small, meaning that the kernel can only represent nearly constant functions. This severely limits its ability to learn complex functions unless the dataset is exponentially large. However, when the model is imbued with ``inductive bias'' (e.g., information about the data generating process), it is able to overcome this challenge and improve model performance significantly. Thus, as the Hilbert space increases, it becomes imperative for QML models to incorporate inductive biases to navigate it effectively. 
\item \textbf{Realizability Bias}\textit{ occurs when a model or algorithm assumes, perhaps implicitly, that all theoretically possible states in a superposition are practically observable, leading to unrealistic or impractical scenarios being considered.} For a system with \textit{n}  qubits, the Hilbert space is \( 2^n \)-dimensional, which means the system can theoretically explore a vast number of states. When a QML model leverages this property, it can consider scenarios that are mathematically possible within the superposition but are not realistically observable, affecting interpretability and model performance. However, this can be mitigated by placing strict constraints on the transition functions and the states accessed by the model.

\end{enumerate}

\subsection*{Measurement Biases}
\begin{enumerate}
     \item \textbf{State-Dependent Bias}\textit{ is a measurement bias where qubits in different states have unequal probabilities of being measured correctly, typically favoring the lower energy state (0) over the higher energy state (1).} This bias arises from qubits' natural tendency to relax to the lower energy state. An experiment by Tannu and Qureshi \cite{tannu2019mitigating} demonstrated this phenomenon and found that the fidelity of an all-zero state was 84\%, but it dropped to 62\% for an all-one state, even though these states are logically interchangeable for many problems (i.e., choice of a variable value as $0$ vs. $1$ is arbitrary). Proposed mitigation strategies for this bias include the ``Invert-And-Measure'' \cite{tannu2019mitigating} method where the higher energy qubits are inverted via an ``X'' gate  and then measured.
    \item \textbf{Sampling Bias}\textit{ arises when the limited number of measurements performed on a quantum system is insufficient to fully capture its state, leading to an incomplete and potentially inaccurate representation of the system.} Quantum measurements project the state of a system onto a basis, collapsing the wavefunction into a specific outcome. To obtain the full probability distribution, one would need to perform a vast number of measurements on different bases. However, in practice, only a limited number of measurements can be performed due to constraints on time, resources, and experimental capabilities. This restricts the sampling to a subspace of the Hilbert space, making it impossible to reconstruct the full probability distribution for higher-order systems. Currently, solutions to this include Quantum State Tomography with Compressed Sensing.\end{enumerate}

\section{Conclusion}
This research examines five biases unique to QML: Encoding, Inductive, Realizability, State-Dependent, and Sampling. These biases, stemming from quantum properties, challenge QML's reliability and performance. 
Our experiment on Encoding bias demonstrates this. We used a QNN with different encodings on the MNIST dataset. The results provide empirical evidence of bias in QML systems, highlighting how different quantum encodings can lead to varying model performances even when the underlying architecture remains constant. While recent advancements in interpretable QML models appear promising, substantial work remains to be done.

\end{document}